\documentclass[sigconf]{acmart}
\AtBeginDocument{%
  }

\copyrightyear{2023}
\acmYear{2023}
\setcopyright{acmlicensed}\acmConference[WWW '23]{Proceedings of the ACM
Web Conference 2023}{May 1--5, 2023}{Austin, TX, USA}
\acmBooktitle{Proceedings of the ACM Web Conference 2023 (WWW '23), May
1--5, 2023, Austin, TX, USA}
\acmPrice{15.00}
\acmDOI{10.1145/3543507.3583446}
\acmISBN{978-1-4503-9416-1/23/04}

\usepackage{multirow}
\usepackage{tikz}
\usepackage{enumitem}
\usepackage{amsmath}

\begin{document}
\settopmatter{printacmref=false}
\settopmatter{printccs=false}
\settopmatter{printfolios=false}
\settopmatter{authorsperrow=4}
\title{NASRec: Weight Sharing Neural Architecture Search for Recommender Systems}

\author{Tunhou Zhang}
\authornote{A majority of this work was done when the first author was an intern at Meta Platforms, Inc.}
\affiliation{%
  \institution{Duke University}
  \city{Durham}
  \country{USA}
 }
 \email{tunhou.zhang@duke.edu}

\author{Dehua Cheng}
\affiliation{%
  \institution{Meta AI}
  \city{Menlo Park}
  \country{USA}
}
\email{dehuacheng@fb.com}

\author{Yuchen He}
\affiliation{%
  \institution{Meta AI}
  \city{Menlo Park}
  \country{USA}
}
\email{yuchenhe@fb.com}

\author{Zhengxing Chen}
\affiliation{%
  \institution{Meta AI}
  \city{Menlo Park}
  \country{USA}
}
\email{czxttkl@fb.com}

\author{Xiaoliang Dai}
\affiliation{%
  \institution{Meta AI}
  \city{Menlo Park}
  \country{USA}
}
\email{xiaoliangdai@fb.com}

\author{Liang Xiong}
\affiliation{%
  \institution{Meta AI}
  \city{Menlo Park}
  \country{USA}
}
\email{lxiong@fb.com}

\author{Feng Yan}
\affiliation{%
  \institution{University of Houston}
  \city{Houston}
  \country{USA}
}
\email{fyan5@central.uh.edu}

\author{Hai Li}
\affiliation{%
  \institution{Duke University}
  \city{Durham}
  \country{USA}
 }
 \email{hai.li@duke.edu}
 
\author{Yiran Chen}
\affiliation{%
  \institution{Duke University}
  \city{Durham}
  \country{USA}
 }
\email{yiran.chen@duke.edu}

\author{Wei Wen}
\authornote{Corresponding author. Intern Manager.}
\affiliation{
  \institution{Meta AI}
  \city{Menlo Park}
  \country{USA}
}
\email{wewen@fb.com}

\begin{abstract}
The rise of deep neural networks offers new opportunities in optimizing recommender systems.
However, 
optimizing recommender systems using deep neural networks requires delicate architecture fabrication.
We propose NASRec, a paradigm that trains a single supernet and efficiently produces abundant models/sub-architectures by weight sharing.
To overcome the data multi-modality and architecture heterogeneity challenges in the recommendation domain, NASRec establishes a large supernet (i.e., search space) to search the full architectures.
The supernet incorporates versatile choice of operators and dense connectivity to minimize human efforts for finding priors.
The scale and heterogeneity in NASRec impose several challenges, such as training inefficiency, operator-imbalance, and degraded rank correlation.
We tackle these challenges by proposing single-operator any-connection sampling, operator-balancing interaction modules, and post-training fine-tuning.
Our crafted models, NASRecNet, show promising results on three Click-Through Rates (CTR) prediction benchmarks, indicating that NASRec outperforms both manually designed models and existing NAS methods with state-of-the-art performance.
Our work is publicly available \href{https://github.com/facebookresearch/NasRec}{here}.

\end{abstract}

\ccsdesc[500]{Information system~Recommender systems}
\ccsdesc[300]{Computing methodologies~Neural networks}
\ccsdesc[100]{Computing methodologies~Discrete space search}

\keywords{recommender systems, neural architecture search, weight sharing, regularized evolution, neural networks}

\maketitle

\section{Introduction}
Deep learning plays an essential role in designing modern recommender systems at web-scale in real-world applications. For example, the most widely used search engines and social medias~\cite{kowald2017temporal,carterette2007evaluating} harness recommender systems (or ranking systems) to optimize the Click-Through Rates (CTR) of personalized pages~\cite{covington2016deep, naumov2019deep, guo2015trustsvd}.
Deep learning models rely on delicate neural architecture engineering.

\begin{table*}[]
\caption{Comparison of NASRec vs. existing NAS methods for recommender systems.}
\vspace{-1em}
\begin{center}
\scalebox{1.0}{
    \begin{tabular}{|c|c|c|c|c|c|}
        \hline
        \multirow{2}{*}{\textbf{Method}} & \textbf{Building} & \textbf{Dense} & \textbf{Full arch} & \textbf{Criteo} 
        & \textbf{Training} \\
         & \textbf{Operators?} & \textbf{Connectivity?} & \textbf{Search?} & \textbf{Log Loss} & \textbf{Cost} \\
    \hline
    \textbf{DNAS}~\cite{krishna2021differentiable} & FC, Dot-Product & \checkmark & & 0.4442 & One supernet \\
    \textbf{PROFIT}~\cite{gao2021progressive} & FC, FM & &  & 0.4427 & One supernet \\
    \textbf{AutoCTR}~\cite{song2020towards} & FC, Dot-Product, FM, EFC & \checkmark & \checkmark & 0.4413 & Many models\\
    \multirow{2}{*}{\textbf{NASRec}} & FC, Gating, Sum, Attention, & \multirow{2}{*}{\checkmark} & \multirow{2}{*}{\checkmark} & \multirow{2}{*}{\textbf{0.4399}} &  \multirow{2}{*}{One supernet} \\ 
     & Dot-Product, FM, EFC & & & & \\
    \hline
    \end{tabular}    
}
\end{center}
\vspace{-1.5em}
\label{tab:intro}
\end{table*}

Deep learning based recommender systems, especially CTR prediction, carries a neural architecture design upon multi-modality features.
In practice, various challenges arise.
The multi-modality features, such as floating-point, integer, and categorical features, present a concrete challenge in feature interaction modeling and neural network optimization.
Finding a good backbone model with heterogeneous architectures assigning appropriate priors upon multi-modality features are common practices in deep learning based recommender systems~\cite{rendle2011fast,richardson2007predicting,he2014practical,cheng2016wide, shan2016deep,guo2017deepfm,lian2018xdeepfm,naumov2019deep}. 
Yet, these approaches still rely on significant manual efforts and suffer from limitations, such as narrow design spaces and insufficient experimental trials bounded by available resources.
As a result, these limitations add difficulty in designing a good feature extractor.

The rise of Automated Machine Learning (AutoML), especially Neural Architecture Search (NAS)~\cite{zoph2018learning,liu2018darts,cai2019once,wen2020neural}, in the vision domain, sheds light in optimizing models of recommender systems.
Weight-Sharing NAS (WS-NAS)~\cite{liu2018darts,cai2019once,bender2020can}  is popularly adopted in vision domain to tackle the design of efficient vision models.
However, applying weight-sharing NAS to recommendation domain is much more challenging than vision domain because of the multi-modality in data and the heterogeneity in architectures. For example,
(1) in vision, inputs of building blocks in ~\cite{cai2019once,wang2020hat} are homogeneous 3D tensors, but recommender systems take in multi-modality features generating 2D and 3D tensors.
(2) Vision models simply stack the same building blocks, and thus state-of-the-art NAS in vision converges to simply searching size configurations instead of architecture motifs, such as channel width, kernel sizes, and layer repeats~\cite{cai2019once,yu2020bignas}. 
However, recommendation models are heterogeneous with each stage of the model using a completely different building block~\cite{cheng2016wide,guo2017deepfm,lian2018xdeepfm,naumov2019deep}.
(3) Vision models mainly use convolutional operator as the main building block while recommender systems are built over heterogeneous operators, such as, Fully-Connected layer, Gating, Sum, Dot-Product, Multi-Head Attention, Factorization Machine, etc.

Due to the aforementioned challenges, study of NAS in recommender systems is very limited.
For example, search spaces in AutoCTR~\cite{song2020towards} and DNAS~\cite{krishna2021differentiable} follow the design principle of human-crafted DLRM~\cite{naumov2019deep} and they only include Fully-Connected layer and Dot-Product as searchable operators. They also heavily reply on manually crafted operators, such as Factorization Machine~\cite{song2020towards} or feature interaction module~\cite{gao2021progressive} in the search space to increase architecture heterogeneity.
Moreover, existing works suffer from either huge computation cost~\cite{song2020towards} or challenging bi-level optimization~\cite{krishna2021differentiable}, and thus they only employ narrow design spaces (sometimes with strong human priors~\cite{gao2021progressive}) to craft architectures, discouraging diversified feature interactions and harming the quality of discovered models.

In this paper, we hereby propose NASRec, a new paradigm to fully enable \textbf{NAS} for \textbf{Rec}ommender systems via Weight Sharing Neural Architecture Search (WS-NAS) under data modality and architecture heterogeneity. Table~\ref{tab:intro} summarizes the advancement of NASRec over other NAS approaches.
We achieve this by first building up a supernet that incorporates much more heterogeneous operators than previous works, including Fully-Connected (FC) layer, Gating, Sum, Dot-Product, Self-Attention, and Embedded Fully-Connected (EFC) layer.
In the supernet, we densely connect a cascade of blocks, each of which includes all operators as options. 
As any block can take in any raw feature embeddings and intermediate tensors by dense connectivity, the supernet is not limited by any particular data modality.
Such supernet design minimizes the encoding of human priors by introducing ``NASRec Search Space'', supporting the nature of data modality and architecture heterogeneity in recommenders, and covering models beyond popular recommendation models
such as Wide \& Deep~\cite{cheng2016wide}, DeepFM~\cite{guo2017deepfm}, DLRM~\cite{naumov2019deep}, AutoCTR~\cite{song2020towards}, DNAS~\cite{krishna2021differentiable}, and PROFIT~\cite{gao2021progressive}.

The supernet essentially forms a search space. We obtain a model by zeroing out some operators and connections in the supernet, that is, a subnet of the supernet is equivalent to a model.
As all subnets share weights from the same supernet, it is dubbed as Weight Sharing NAS.
To efficiently search models/subnets in the NASRec search space, we advance one-shot approaches~\cite{cai2019once,yu2020bignas} to recommendation domain.
We propose \textit{Single-operator Any-connection sampling} to decouple operator selections and increase connection coverage,
\textit{operator-balancing interaction} blocks to fairly train subnets in the supernet, and \textit{post-training fine-tuning} to reduce weight co-adaptation.
These approaches enable a better training efficiency and ranking of subnet models in the supernet, resulting in $\sim$0.001 log loss reduction of searched models on full NASRec search space.

We evaluate our NAS-crafted models, NASRecNets on three popular CTR benchmarks and demonstrate significant improvements compared to both hand-crafted models and NAS-crafted models.
Remarkably, NASRecNet advances the state-of-the-art with log loss reduction of $\sim 0.001$, $\sim 0.003$ on Criteo and KDD Cup 2012, respectively.
On Avazu, NASRec advances the state-of-the-art PROFIT~\cite{gao2021progressive} with AUC improvement of $\sim 0.002$ and on-par log loss, while outperforming PROFIT~\cite{gao2021progressive} on Criteo by $\sim$0.003 log loss reduction.

NASRec only needs to train a single supernet thanks to the efficient weight sharing mechanism, and thus greatly reduces the search cost. We summarize our major contributions below.
\begin{itemize}[noitemsep,leftmargin=*]
    \item We propose NASRec, a new paradigm to scale up automated modeling of recommender systems.
    NASRec establishes a flexible supernet (search space) with minimal human priors, overcoming data modality and architecture heterogeneity challenges in the recommendation domain.
    \item We advance weight sharing NAS to recommendation domain by introducing single-operator any-connection sampling, operator-balancing interaction modules, and post-training fine-tuning.
    \item Our crafted models, NASRecNet, outperforms both hand-crafted models and NAS-crafted models with a smaller search cost.
\end{itemize}

\section{Related Work}

\noindent \textbf{Deep learning based recommender systems.}
Machine-based recommender systems such as Click-Through Rates (CTR) prediction has been thoroughly investigated in various approaches, such as Logistic Regression~\cite{richardson2007predicting}, and Gradient-Boosting Decision Trees~\cite{he2014practical}.
More recent approaches study deep learning based interaction of different types of features via  Wide \& Deep Neural Networks~\cite{cheng2016wide}, DeepCrossing~\cite{shan2016deep}, Factorization Machines~\cite{guo2017deepfm,lian2018xdeepfm}, Dot-Product~\cite{naumov2019deep} and gating mechanism~\cite{wang2017deep, wang2021dcn}.
Another line of research seeks efficient feature interactions, such as feature-wise multiplications~\cite{wang2021masknet} and sparsifications~\cite{deng2021deeplight} to build light-weight recommender systems.
Yet, these works operate the cost of tremendous manual efforts and suffer from sub-optimal performance and constrained design choices due to the limitations in resource supply.
Our work establishes a new paradigm on learning effective recommender models by crafting a scalable "NASRec search space" that incorporates all popular design motifs in existing works.
The new NASRec search space supports a wide range of design choices and enables scalable optimization to craft recommender models of varying requirements. 

\noindent \textbf{Neural Architecture Search.}     
Neural Architecture Search automates the design of Deep Neural Networks in various applications: the popularity of Neural Architecture Search is consistently growing in brewing Computer Vision~\cite{zoph2018learning,liu2018darts,wen2020neural,cai2019once}, Natural Language Processing~\cite{so2019evolved,wang2020hat}, and Recommendation Systems~\cite{song2020towards,gao2021progressive,krishna2021differentiable}.
Recently, Weight-Sharing NAS (WS-NAS)~\cite{cai2019once,wang2020hat} attracts the attention of researchers: it trains a supernet that represents the whole search space directly on target tasks, and efficiently evaluate subnets (i.e., sub-architectures of supernet) with shared supernet weights.
Yet, carrying WS-NAS on recommender systems is challenging because recommender systems are brewed upon heterogeneous architectures that are dedicated to interacting multi-modality data, thus require more flexible search spaces and effective supernet training algorithms. 
Those challenges induce the co-adaptation problem~\cite{bender2018understanding} and operator-imbalance problem~\cite{liang2019darts+} in WS-NAS, providing a lower rank correlation to distinguish models.
NASRec addresses them by proposing single-operator any-connection sampling, operator-balancing interaction modules, and post-training fine-tuning.
\section{Hierarchical NASRec Space for Recommender Systems}
\begin{figure*}[t]
    \begin{center}
    \includegraphics[width=0.8\linewidth]{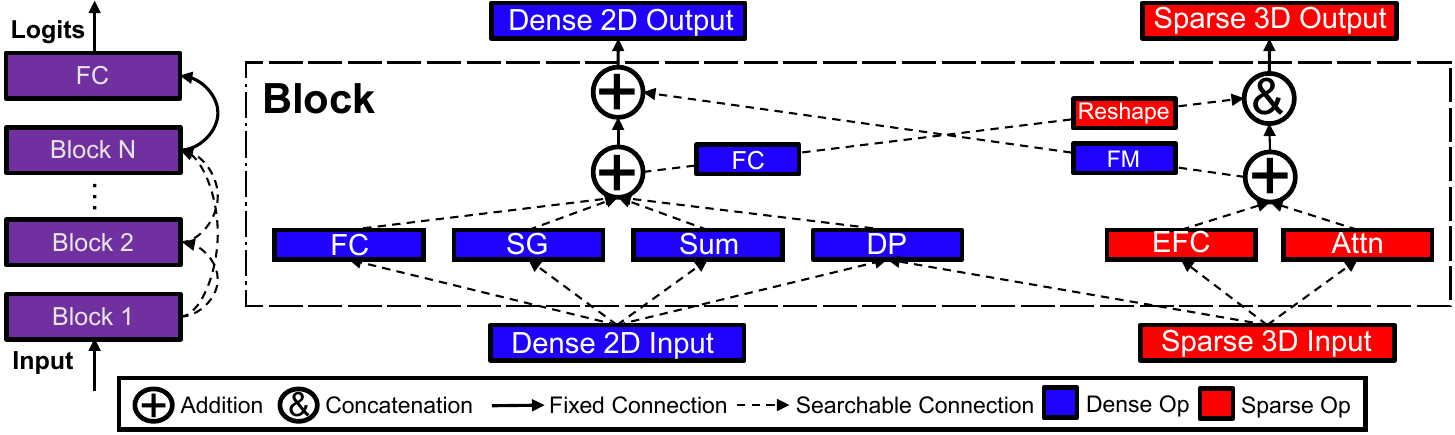}
    \vspace{-1em}
    \caption{Overview of NASRec search space. NASRec search space enables a full architecture search on building operators and dense connectivity. Here, \textcolor{blue}{``blue''} blocks produce a dense output, and \textcolor{red}{``red''} blocks produce a sparse output.}
    \label{fig:wsnas_search_space}    
    \end{center}
    \vspace{-1em}
\end{figure*}

To support data modality and architecture heterogeneity in recommender systems, the flexibility of search space is the key.
We establish a new paradigm free of human priors by introducing \textit{NASRec search space}, a hierarchical search space design that incorporates heterogeneous building operators and dense connectivity, see Figure \ref{fig:wsnas_search_space}.
The major manual process in designing the search space is simply collecting common operators used in existing approaches~\cite{guo2017deepfm,lian2018xdeepfm, naumov2019deep, song2020towards,wang2017deep, wang2021dcn}.
Beyond that, we further incorporate the prevailing Transformer Encoder~\cite{vaswani2017attention} into the NASRec search space for better flexibility and higher potential in searched architectures, thanks to its dominance in applications such as ViT~\cite{dosovitskiy2020image} for image recognition, Transformer~\cite{vaswani2017attention} for natural language processing, and its emerging exploration in recommender systems~\cite{chen2019behavior, gao2020modularized}.

Next, we demonstrate the NASRec search space.

\subsection{NASRec Search Space}
\label{sec:search_space}
In recommender systems, we define a dense input as $X_{d} \in \mathbb{R}^{B \times dim_d}$ which is a 2D tensor from either raw dense features or generated by operators, such as FC, Gating, Sum, and Dot-Product.
A sparse input $X_{s} \in \mathbb{R}^{B \times N_s \times dim_s}$ is a 3D tensor of sparse embeddings either generated by raw sparse/categorical features or by operators such as EFC and self-attention.
Similarly, a dense or sparse output (i.e., $Y_{d}$ or $Y_{s}$) is respectively defined as a 2D or 3D tensor produced via a corresponding building blocks/operators.
In NASRec, all sparse inputs and outputs share the same $dim_s$, which equals to the dimension of raw sparse embeddings.
Accordingly, we define a dense (sparse) operator as an operator that produces a dense (sparse) output. 
In NASRec, dense operators include FC, Gating, Sum, and Dot-Product which form the ``dense branch'' (marked in \textcolor{blue}{blue}), and sparse operators include EFC and self-attention, which form the ``sparse branch'' (marked in \textcolor{red}{red}). 

A candidate architecture in NASRec search space is a stack of $N$ choice blocks, followed by a final FC layer to compute logit.
Each choice block admits an arbitrary number of multi-modality inputs, each of which is $X=(X_{d}, X_{s})$ from a previous block or raw inputs, and produces a multi-modality output $Y=(Y_{d}, Y_{s})$ of both a dense tensor $Y_{d}$ and a sparse tensor $Y_{s}$ via internal building operators.
Within each choice block, we can sample operators for search.

We construct a supernet to represent the NASRec search space, see Figure \ref{fig:wsnas_search_space}.
The supernet subsumes all possible candidate models/subnets and performs weight sharing among subnets to simultaneously train all of them.
We formally define the NASRec supernet $\mathcal{S}$ as a tuple of connections $\mathcal{C}$, operators $\mathcal{O}$, and dimensions $\mathcal{D}$ as follows: $\mathcal{S} = (\mathcal{C}, \mathcal{D}, \mathcal{O})$ over all $N$ choice blocks. Specifically, the operators: $\mathcal{O}=[O^{(1)}, ..., O^{(N)}]$ enumerates the set of building operators from choice block $1$ to $N$.
The connections: $\mathcal{C}=[C^{(1)}, ..., C^{(N)}]$ contains the connectivity $<i, j>$ between choice block $i$ and choice block $j$.
The dimension: $\mathcal{D}=[D^{(1)}, ..., D^{(N)}]$ contains the dimension settings from choice block $1$ to $N$.

A subnet $S_{sample}=(\mathcal{O}_{sample}, \mathcal{C}_{sample}, \mathcal{D}_{sample})$ in the supernet $\mathcal{S}$ represents a model in NASRec search space. 
A block uses addition to aggregate the outputs of sampled operators in each branch (i.e. ``dense branch'' or ``sparse branch''). When the operator output dimensions do not match, we apply a zero masking to mask out the extra dimension.
A block uses concatenation $Concat$
to aggregate the outputs from sampled connections.
Given a sampled subnet $S_{sample}$, the input $X^{(N)}$ to choice block $N$ is computed as follows given a list of previous block outputs $\{Y^{(1)}, ..., Y^{(N-1)}\}$ and the sampled connections $C_{sample}^{(N)}$:

\begin{equation}
\small
    \vspace{-1em}
    X_d^{(N)} = Concat_{i=1}^{N-1}[Y_d^{(i)} \cdot \mathbf{1}_{<i, N> \in C^{(N)}_{sample}}],    
\end{equation}
\begin{equation}
\small
    \vspace{-0.5em}
    X_s^{(N)} = Concat_{i=1}^{N-1}[Y_s^{(i)} \cdot \mathbf{1}_{<i, N> \in C^{(N)}_{sample}}].
\end{equation}

Here, $\mathbf{1}_b$ is 1 when $b$ is true otherwise 0.

A building operator $o \in O^{(N)}_{sample}$ transforms the concatenated input $X^{(N)}$ into an intermediate output with a sampled dimension $D^{(N)}_{sample}$. This is achieved by a mask function that applied on the last dimension for dense output and middle dimension for sparse output.
For example, a dense output $Y^{(N)}_{d}$ is obtained as follows:

\begin{equation}
\small
    Y_d^{(N)} = \sum_{o \in \mathcal{O}} {\mathbf{1}_{o \in \mathcal{O}_{sample}^{(N)}} \cdot Mask(o(X_d^{(N)}), D^{(N)}_{sample, o})}.
\end{equation}
where
\begin{equation}
\label{eq:mask_dim}
\small
    Mask(V, d) = \begin{cases}
V_{:, i}, \text{ if } i<d \\
0, \text{ Otherwise.}\\
\end{cases}.
\end{equation}

Next, we clarify the set of building operators as follows:
\begin{itemize}[noitemsep,leftmargin=*]
    \item \textbf{Fully-Connected (FC) layer.} Fully-Connected layer is the backbone of DNN models for recommender systems~\cite{cheng2016wide} that extracts dense representations. FC is applied on 2D dense inputs, and followed by a ReLU activation. 
    
    \item \textbf{Sigmoid Gating (SG) layer.} 
    We follow the intuition in ~\cite{wang2021dcn,chen2019behavior} and employ a dense building operator, Sigmoid Gating, to enhance the potential of the search space. 
    Given two dense inputs $X_{d1}\in \mathbb{R}^{B \times dim_{d1}}$ and $X_{d2}\in \mathbb{R}^{B \times dim_{d2}}$, Sigmoid Gating interacts these two inputs as follows: $SG(X_{d1}, X_{d2}) = sigmoid(FC(X_{d1})) * X_{d2}$.
    If the dimension of two dense inputs does not match, a zero padding is applied on the input with a lower dimension.
        
    \item \textbf{Sum layer.} This dense building operator adds two dense inputs: $X_{d1} \in \mathbb{R}^{B \times dim_{d1}}$, $X_{d2} \in \mathbb{R}^{B \times dim_{d2}}$ and merges two features from different levels of the recommender system models by simply performing $Sum(X_{d1}, X_{d2})=X_{d1} + X_{d2}$.
    Similar to Sigmoid Gating, a zero padding is applied on the input with a lower dimension.

    \item \textbf{Dot-Product (DP) layer.} We leverage Dot-Product to grasp the interactions among multi-modality inputs via a pairwise inner products.
    Dot-Product can take dense and/or sparse inputs, and produce a dense output.
    These sparse inputs, after being sent to ``dense branch'', can later take advantage of the dense operators to learn better representations and interactions.
    Given a dense input $X_{d} \in \mathbb{R}^{B \times dim_{d}}$ and a sparse input $X_{s} \in \mathbb{R}^{B \times N_c \times dim_{s}}$, a Dot-Product first concatenate them as $X = Concat[X_d, X_s]$, and then performs pair-wise inner products: $DP(X_d, X_s)=Triu(XX^{T})$. $dim_{d}$ is first projected to $dim_{s}$ if they do not match.

    \item \textbf{Embedded Fully-Connected (EFC) layer}. An EFC layer is a sparse building operator that applies FC along the middle dimension. Specifically, an EFC with weights $W \in \mathbb{R}^{N_{in} \times N_{out}}$ transforms an input $X_s \in \mathbb{R}^{B \times N_{in} \times dim_{s}}$ to $Y_s \in \mathbb{R}^{B \times N_{out} \times dim_{s}}$
    
    \item \textbf{Attention (Attn) layer.} Attention layer is a sparse building operator that utilizes Multi-Head Attention (MHA) mechanism to learn the weighting of sparse inputs and better exploit their interaction in recommendation systems. Here, We apply Transformer Encoder on a given sparse input $X_{s} \in \mathbb{R}^{B\times N_s \times dim_{s}}$, with identical queries, keys, and values.
\end{itemize}

We observe that the aforementioned set of building operators provide  opportunities for the sparse inputs to transform into the ``dense branch''.
Yet, these operators do not permit a transformation of dense inputs towards the ``sparse branch''.
To address this limitation,
we introduce "\textbf{dense-sparse merger}"
allow dense/sparse outputs to optionally merge into the ``sparse/dense branch''. Dense-sparse merger contains two major components.
\begin{itemize}[noitemsep,leftmargin=*]
    \item "Dense-to-sparse" merger. This merger first projects the dense outputs $X_{d}$ using a FC layer, then uses a reshape layer to reshape the projection into a 3D sparse tensor. The reshaped 3D tensor is merged into the sparse output via concatenation. 
    \item "Sparse-to-dense" merger. This merger employs a Factorization Machine (FM)~\cite{guo2017deepfm} to convert the sparse output into a dense representation, then add the dense representation to dense output. 
\end{itemize}

Beyond the rich choices of building operators and mergers, each choice block can also receive inputs from any preceding choice blocks, and raw input features. 
This involves an exploration of any connectivity among choice blocks and raw inputs, extending the wiring heterogeneity for search.

\subsection{Search Components} 
In NASRec search space, we search the connectivity, operator dimensions, and building operators in each choice block.
We illustrate the three key search components as follows:

\begin{itemize}[noitemsep,leftmargin=*]
\item \textbf{Connection.}
We place no restrictions on the number of connections that a choice block can receive: each block can choose inputs from an arbitrary number of preceding blocks and raw inputs. 
Specifically, the n-th choice block can connect to any previous $n-1$ choice blocks and the raw dense (sparse) features.
The outputs from all preceding blocks are concatenated as inputs for dense (sparse) building blocks.
We separately concatenate the dense (sparse) outputs from preceding blocks. 

\item \noindent \textbf{Dimension.}
In a choice block, different operators may produce different tensor dimensions.
In NASRec, we set the output sizes of FC and EFC to $dim_{d}$ and $N_{s}$, respectively; and other operator outputs in dense (sparse) branch are linearly projected to $dim_{d}$ ($N_{s}$).
This ensures operator outputs in each branch have the same dimension and can add together. This also give the maximum dimensions $dim_{d}$ and $N_{s}$ for the dense output $Y_{d} \in \mathbb{R}^{B \times dim_d}$ and the sparse output $Y_{s} \in \mathbb{R}^{B \times N_s \times dim_s}$.
Given a dense or sparse output, a mask in Eq.~\ref{eq:mask_dim} zeros out the extra dimensions, which allows flexible selections of dimensions of building operators.

\item \textbf{Operator.} Each block can choose at least one dense (sparse) building operator to transform inputs to a dense (sparse) output. Each block should maintain at least one operator in the dense (sparse) branch to ensure the flow of information from inputs to logit.
We independently sample building operators in the dense (sparse) branch to form a validate candidate architecture.
In addition, we independently sample dense-sparse mergers to allow optional dense-to-sparse interaction.

\end{itemize}

We craft two NASRec search spaces as examples to demonstrate the power of NASRec search space.

\begin{itemize}[noitemsep,leftmargin=*]
    \item \textit{NASRec-Small.} We limit the choice of operators within each block to FC, EFC, and Dot-Product, and allow any connectivity between blocks. This provides a similar scale of search space as AutoCTR~\cite{song2020towards}.   

    \item \textit{NASRec-Full.} We enable all building operators, mergers and connections to construct an aggressive search space for exploration with minimal human priors. Under the constraint that at least one operator must be sampled in both dense and sparse branch, the \textit{NASRec-Full} search space size is $15^{N}\times$ of \textit{NASRec-Small}, where $N$ is the number of choice blocks. This full search space extremely tests the capability of NASRec.
    \end{itemize}

The combination of full dense connectivity search and independent dense/sparse dimension configuration gives the NASRec search space a large cardinality.
\textit{NASRec-Full} has $N=7$ blocks, containing up to $5\times 10^{33}$ architectures with strong heterogeneity.
With minimal human priors and such unconstrained search space, brutal-force sample-based methods may take enormous time to find a state-of-the-art model.
\section{Weight sharing Neural Architecture Search for Recommender Systems}
A NASRec supernet simultaneously brews different subnet models in the NASRec search space, yet imposes challenges to training efficiency and ranking quality due to its large cardinality.
In this section, we first propose a novel path sampling strategy, \textit{Single-operator Any-connection} sampling, that decouples operator sampling with a good connection sampling converge.
We further observe the operator imbalance phenomenon induced by some over-parameterized operators, and tackle this issue by \textit{operator-balancing interaction} to improve supernet ranking.
Finally, we employ \textit{post-training fine-tuning} to alleviate weight co-adaptation, and further utilize regularized evolution to obtain the best subnet.
We also provide a set of insights that effectively explore the best recommender models.

\begin{figure}[t]
    \begin{center}
    \includegraphics[width=1.0\linewidth]{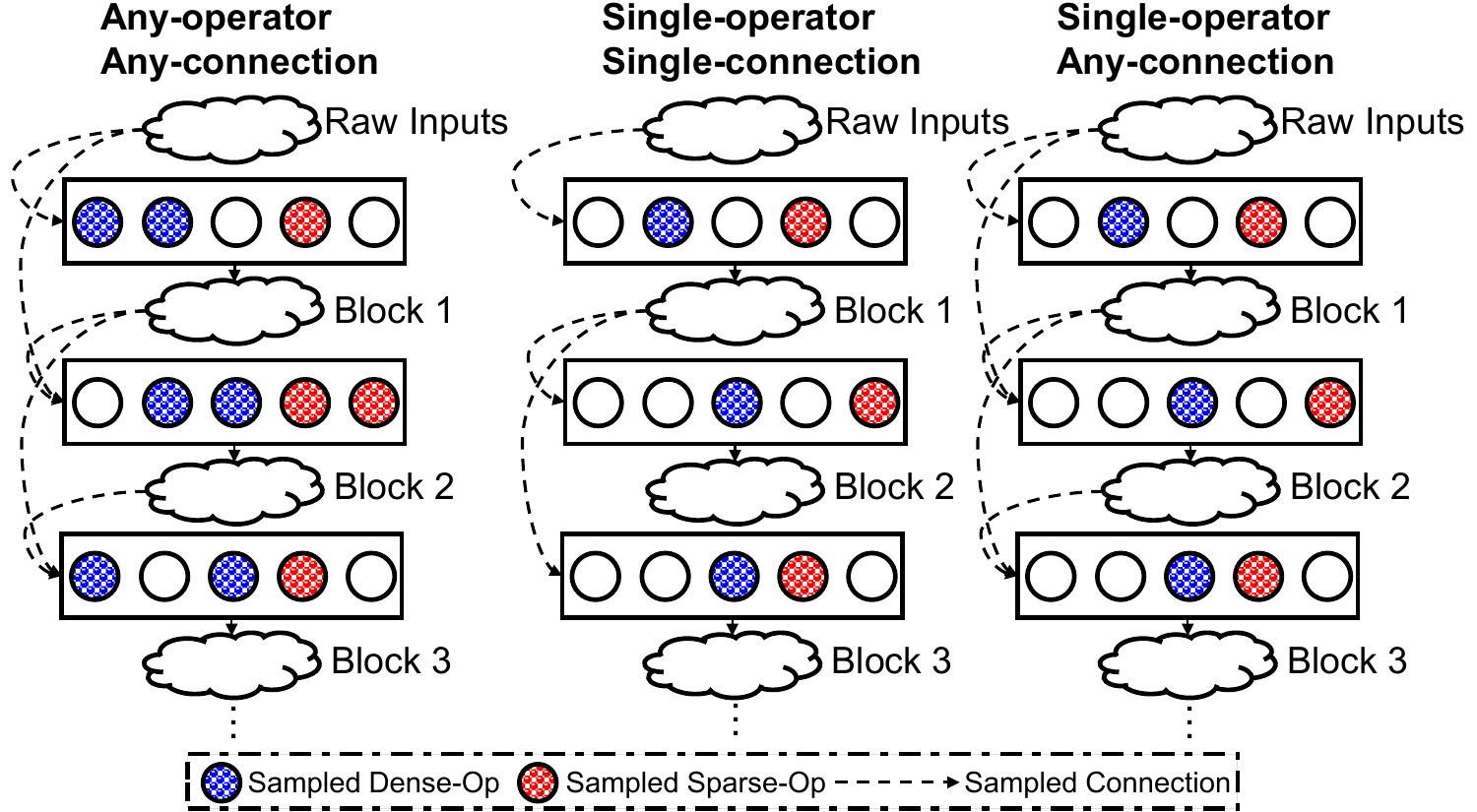}
    \vspace{-2em}
    \caption{We propose Single-operator Any-connection path sampling by combining the advantages of the first two sampling strategies. 
    Here, dashed connections and operators denotes a sampled path in supernet.}
    \label{fig:single_path}    
    \end{center}
    \vspace{-2em}
\end{figure}

\subsection{Single-operator Any-Connection Sampling}
\label{sec:single_path}
The supernet training adopts a drop-out like approach.
At each mini-batch, we sample and train a subnet. During training, we train lots of subnets under weight sharing, with the goal that subnets are well trained to predict the performance of models.
Sampling strategies are important to meet the goal.
We explore three path sampling strategies depicted in Figure \ref{fig:single_path} and discover Single-operator Any-Connection sampling is the most effective way:
\begin{itemize}[noitemsep,leftmargin=*]
    \item \textbf{Single-operator Single-connection strategy.} This path sampling strategy has its root in Computer Vision~\cite{guo2020single}: it uniformly samples a single dense and a single sparse operator in each choice block, and uniformly samples a single connection as an input to a block.
    The strategy is efficient because, on average, only a small subnet is trained at one mini-batch, however, this strategy only encourages chain-like formulation of models without extra connectivity patterns.
    The lack of connectivity coverage yields slower convergence, poor performance, and inaccurate ranking of models as we will show.
    \item \textbf{Any-operator Any-connection Strategy.} This sampling strategy increases the coverage of sub-architectures of supernet during subnet training: it uniformly samples an arbitrary number of dense and sparse operators in each choice block, and uniformly sample an arbitrary number of connections to aggregate different block outputs.
    Yet, the training efficiency is poor when training sampled large subnets. 
    More importantly, the weight co-adaptation of multiple operators within a choice block may affect independent evaluation of the subnets, and thus eventually lead to poor ranking quality as we will show.
    \item \textbf{Single-operator Any-connection}.
    We propose this path sampling strategy to combine the strengths from above two strategies.
    Single-operator Any-connection samples a single dense and a single sparse operator in each choice block, and samples an arbitrary number of connections to aggregate the outputs from different choice blocks.
    The key insight of this strategy is separating the sampling of parametric operators to avoid the co-adaptation of weights, and allowing arbitrarily sample of non-parametric connections to gain a good coverage of the NASRec search space.
\end{itemize}

\begin{figure}[t]
    \begin{center}
    \includegraphics[width=0.9\linewidth]{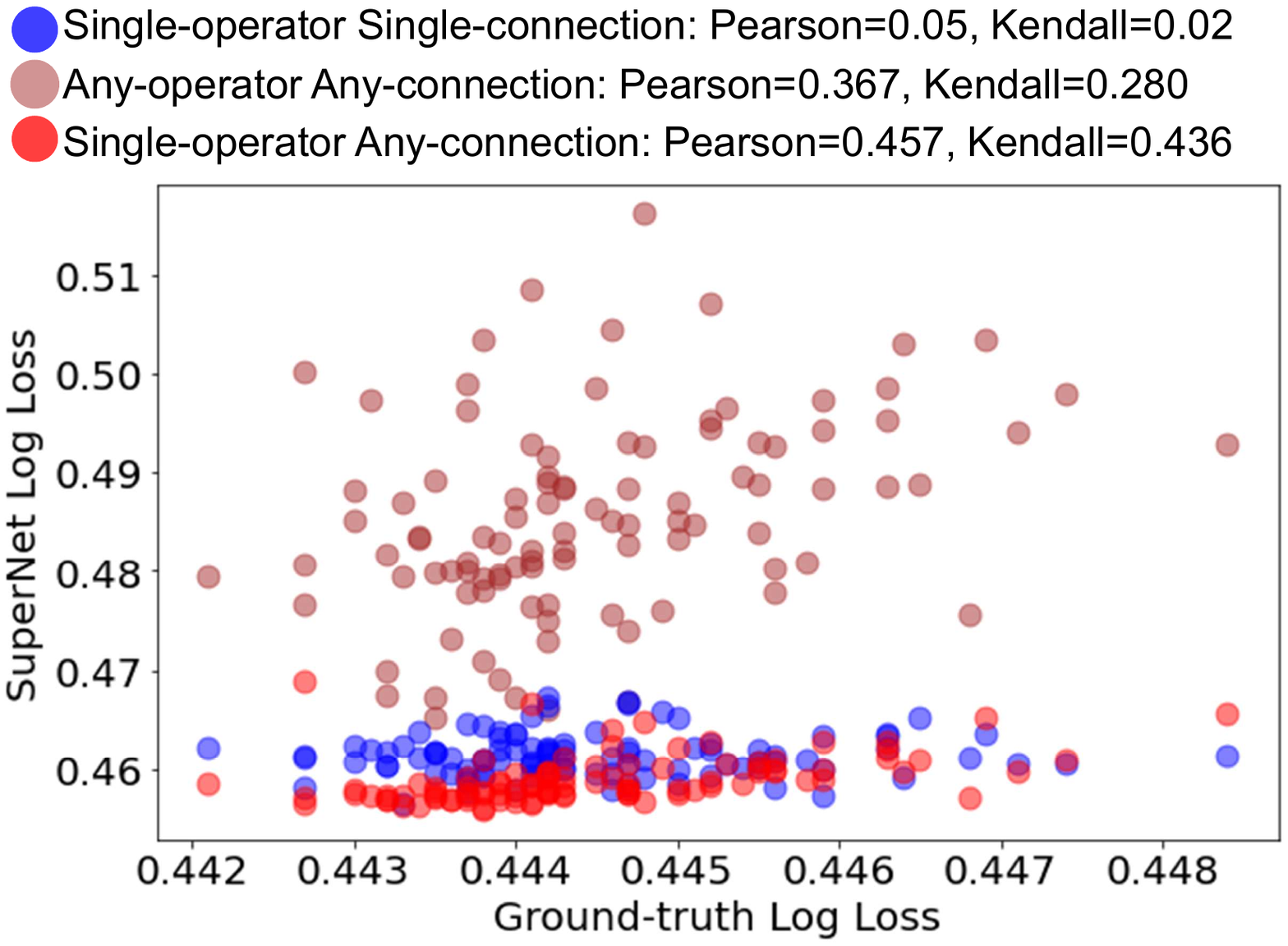}
    \vspace{-1em}
    \caption{Ranking evaluation of various path sampling strategies on \textit{NASRec-Full} supernet. We evaluate all ranking coefficients over 100 randomly sampled subnets on Criteo.}
    \label{fig:single_path_evaluation}    
    \end{center}
    \vspace{-2em}
\end{figure}

Compared to Any-operator Any-connection sampling, 
single-operator Any-connection sampling achieves higher training efficiency: the reduced number of sampled operators reduces the training cost by up to 1.5$\times$.
In addition, Single-operator Any-connection samples medium-sized networks more frequently.
These medium-sized networks achieve the best trade-off between model size and performance as we will show in Table ~\ref{tab:model_complexity}.

We evaluate the ranking of subnets by WS-NAS on Criteo and by 100 randomly sampled networks in Figure \ref{fig:single_path_evaluation}.
Here, we adopt the design of operator-balancing interaction modules in Section 4.2 to maximize the potential of each path sampling strategy.
In the figure, the y-axis is the Log Loss of subnets, whose weights are copied from corresponding architectures in the trained supernet.
Single-operator Any-connection achieves at least 0.09 higher Pearson's Rho and 0.15 higher Kendall's Tau compared to other path sampling strategies.
In addition, we observe that Single-operator Any-connection sampling allows better convergence of the NASRec supernet and subnets that inherit weights from supernet achieve lower log loss during validation, leading to a better exploitation of their ground-truth performance for a better ranking quality.

\begin{figure}[t]
    \begin{center}
    \includegraphics[width=0.9\linewidth]{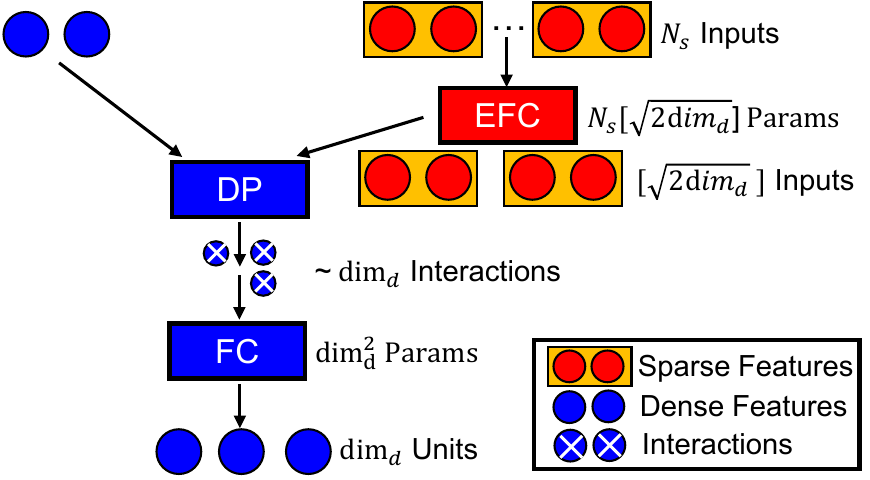}
    \vspace{-1em}
    \caption{Operator-balancing interaction inserts a simple EFC layer before Dot-Product to ensure linear parameter consumption and balance building operators.}
    \label{fig:op_imbalance}
    \vspace{-2em}
    \end{center}
\end{figure}

\subsection{Operator-Balancing Interaction Modules}
Recommender systems involve multi-modality data with an indefinite number of inputs, for example, a large number of sparse inputs.
We define operator imbalance as the imbalance of the numbers of weights between operators within a block.
In weight-sharing NAS, operator imbalance may cause the issue that supernet training may favor operators with more weights. This will offset the gains due to poor ranking correlations of subnets: the subnet performance in supernet may deviate from its ground-truth performance when trained from scratch.
We identify that, in our NASRec, such an issue is strongly related to the Dot-Product operator, and provide mitigation to address such operator imbalance.

Given $N_s$ sparse embeddings, a Dot-Product block produces $N_s^2/2$ pairwise interactions as a quadratic function on the number of sparse embeddings.
As detailed in Section \ref{sec:search_space}, the supernet requires a linear projection layer (i.e., FC) to match the output dimensions of operators within each choice block.
Typically for Dot-Product, this leads to an extra $(N_s^2 \cdot dim_{d} /2)$ trainable weights.

However, the weight consumption of such projection layer is large given a large number of sparse embeddings.
For example, given $N_s=448$ and $dim_{d}=512$ in a 7-block NASRec supernet, the projection layer induces over $50M$ parameters in the NASRec supernet, which has a similar scale of parameter consumption with sparse embedding layers.
Such tremendous weight parameterization is a quadratic function of the number of sparse inputs $N_s$, yet other building operators have much fewer weights, such as, the number of trainable weights in EFC is a linear function of the number of sparse inputs $N_s$.
As a result, the over-parameterization in Dot-Product leads to an increased convergence rate for the Dot-Product operator and consequently favor parameter-consuming subnets with a high concentration of Dot-Product operations as we observed.
In addition, the ignorance of other heterogeneous operators other than Dot-Product provides a poor ranking of subnets, leading to sub-optimal performance on recommender systems.

We insert a simple EFC as a projection layer before the Dot-Product to mitigate such over-parameterization, see Figure \ref{fig:op_imbalance}.
Our intuition is projecting the number of sparse embeddings in Dot-Product to $[\sqrt{2dim_{d}}]$, such that the following Dot-Product operator produces approximately $dim_{d}$ outputs that later requires a minimal projection layer to match the dimension.
As such, the Dot-Product operator consumes at most $(dim_{d}^2+N_s[\sqrt{2dim_{d}}])$ trainable weights and ensures a linear growth of parameter consumption with the number of sparse EFC $N_s$.
Thus, we balance interaction operator to allow a more similar convergence rate of all building operators.
Table \ref{tab:op_imbalance_results} reflects a significant enhancement on the training efficiency and ranking quality of the \textit{NASRec-Full} supernet with Single-operator Any-connection path sampling strategy.

\begin{table}[t]
    \begin{center}
    \caption{Operator-Balancing Interactions reduce supernet training cost and improve ranking of subnets.}
    \vspace{-1em}
    \scalebox{0.85}{
    \begin{tabular}{|c|c|c|c|}
    \hline
         \textbf{Interaction Type} & \textbf{Training Cost} & \textbf{Pearson's} $\rho$ & \textbf{Kendall's $\tau$} \\
    \hline
    Imbalanced DP & \large 4 Hours &  \large  0.31 & \large  0.32 \\
    Balanced DP & \large 1.5 Hours & \large  0.46 & \large  0.43 \\ 
    \hline
    \end{tabular}        
    }
    \label{tab:op_imbalance_results}
    \vspace{-2em}
    \end{center}
\end{table}

\label{sec:path_ft}
\subsection{Post-training Fine-tuning}
Although dropout-like subnet training provide a great way to reduce the adaptation of weights for a specific subnet,
the subnet performance prediction by supernet can fail when weights should not share across some subnets.
After the supernet training and during a stand alone subnet evaluation, we carry a post-training fine-tuning that re-adapt its weights back to the specific subnet. This can re-calibrate the weights which are corrupted when training other subnets during the supernet training.
In practice, we find that fine-tuning the last FC on the target dataset for a few training steps (e.g., 0.5K) is good enough.
With only marginal additional search cost, this novel post-training fine-tuning technique boosts the ranking of subnets by addressing the underlying weight adaptation issue, and thus provides a better chance to discover better models for recommender systems.

\begin{table}[b]
\begin{center}
\vspace{-1.5em}
\caption{Effects of post-training fine-tuning on different path sampling strategies on \textit{NASRec-Full}. We demonstrate Pearson's $\rho$ and Kendall's $\tau$ over 100 random subnets on Criteo.}
\vspace{-1em}
\scalebox{0.7}{
    \begin{tabular}{|c|cc|cc|}
    \hline
    \multirow{2}{*}{\textbf{Path Sampling Strategy}} & \multicolumn{2}{|c|}{\textbf{No Fine-tuning}} &
    \multicolumn{2}{|c|}{\textbf{Fine-tuning}}  \\
    & Pearson's $\rho$  & Kendall's $\tau$ & Pearson's $\rho$  & Kendall's $\tau$ \\
    \hline
    Any-operator Any-connection & \large 0.37 & \large 0.28 & \large  0.46 & \large  0.43 \\
    Single-operator Single-connection & \large  0.05 & \large  0.02 & \large  0.43 & \large  0.29 \\
    Single-operator Any-connection & \large  0.46 & \large  \large  0.43 & \large  0.57 & \large  0.43 \\
    \hline
    \end{tabular}
}
    \label{tab:ft_subnets}    
\end{center}
\end{table}

Table \ref{tab:ft_subnets} demonstrates the improvement of post-training fine-tuning on different path sampling strategies.
Surprisingly, post-training fine-tuning achieves decent ranking quality improvement under Single-operator Single-connection  and Any-operator Any-connection path sampling strategy.
This is because subnets under these strategies do not usually converge well in supernet: they either suffer from poor supernet coverage, or poor convergence induced by co-adaptation.
The fine-tuning process releases their potential and approaches their real performance on the target dataset.
Remarkably, Single-operator Any-connection path sampling strategy cooperates well with post-training fine-tuning, and achieves the global optimal Pearson's $\rho$ and Kendall's $\tau$ ranking correlation among different approaches, with at least $0.14$  Pearson's $\rho$ and Kendall's $\tau$ improvement on \textit{NASRec-Full} search space over Single-operator Single-connection sampling with fine-tuning.

\subsection{Evolutionary Search on Best Models}
We utilize regularized evolution~\cite{real2019regularized} to obtain the best child subnet in NASRec search space, including \textit{NASRec Small} and \textit{NASRec-Full}. Here, we first introduce a single mutation of a hierarchical genotype with the following sequence of actions in one of the choice blocks:
\begin{itemize}[noitemsep,leftmargin=*]
\item Re-sample the dimension of one dense building operator.
\item Re-sample the dimension of one sparse building operator.
\item Re-sample one dense building operator.
\item Re-sample one sparse building operator.
\item Re-sample its connection to other choice blocks.
\item Re-sample the choice of dense-to-sparse/sparse-to-dense merger that enables the communication between dense/sparse outputs.
\end{itemize}

\begin{table*}[t]
    \begin{center}
    \caption{Performance of NASRec on General CTR Predictions Tasks.}
    \vspace{-1em}
    \scalebox{0.98}{
    \begin{tabular}{|c|c|cc|cc|cc|c|c|}
    \hline
     & \multirow{2}{*}{\textbf{Method}} & \multicolumn{2}{|c|}{\textbf{Criteo}}  &  \multicolumn{2}{|c|}{\textbf{Avazu}} & \multicolumn{2}{|c|}{\textbf{KDD Cup 2012}}  & \textbf{Search Cost} \\
      & &  Log Loss & AUC & Log Loss & AUC & Log Loss & AUC & (GPU days) \\
    \hline \hline
    \multirow{4}{*}{\textbf{Hand-crafted Arts}} & DLRM~\cite{naumov2019deep} & 0.4436 & 0.8085 & 0.3814 & 0.7766 & 0.1523 & 0.8004 & - \\
    & xDeepFM~\cite{lian2018xdeepfm} & 0.4418 & 0.8052 & - & - & - & - & - \\
    & AutoInt+~\cite{song2019autoint} & 0.4427 & 0.8090 & 0.3813 & 0.7772 & 0.1523 & 0.8002 & - \\
    & DeepFM~\cite{guo2017deepfm} & 0.4432 & 0.8086 & 0.3816 & 0.7767 & 0.1529 & 0.7974 & -\\
    \hline

    \multirow{7}{*}{\textbf{NAS-crafted Arts}} & DNAS~\cite{krishna2021differentiable} & 0.4442 & - & - & - & - & - & - \\
    & PROFIT~\cite{gao2021progressive} & 0.4427 & 0.8095 & \textbf{0.3735} & 0.7883 & - & - & $\sim$0.5 \\
    & AutoCTR~\cite{song2020towards} & 0.4413 & 0.8104 & 0.3800 & 0.7791 & 0.1520 & 0.8011 & $\sim$0.75 \\
    & Random Search @ \textit{NASRec-Small} & 0.4411 & 0.8105 & 0.3748 & 0.7885 & 0.1500 & 0.8123 & 1.0 \\
    & Random Search @ \textit{NASRec-Full} & 0.4418 & 0.8098 & 0.3767 & 0.7853 & 0.1509 & 0.8071 & 1.0 \\
    & NASRecNet @ \textit{NASRec-Small} & \textbf{0.4399} &\textbf{0.8118} & 0.3747 & 0.7887 & \textbf{0.1495} & \textbf{0.8135} & $\sim$0.25 \\
    & NASRecNet @ \textit{NASRec-Full} & \textbf{0.4408} & \textbf{0.8107} & \textbf{0.3737} & \textbf{0.7903} & \textbf{0.1491} & \textbf{0.8154} & $\sim$0.3 \\
    \hline
    \end{tabular}
    \label{tab:ctr_results}
    }
    \end{center}
    \vspace{-1em}
\end{table*}

\section{Experiments}

We first show the detailed configuration that NASRec employs during architecture search, model selection and final evaluation.
Then, we demonstrate empirical evaluations on three popular recommender system benchmarks for Click-Through Rates (CTR) prediction: Criteo\footnote{\hyperlink{https://www.kaggle.com/c/criteo-display-ad-challenge}{https://www.kaggle.com/c/criteo-display-ad-challenge}}, Avazu\footnote{\hyperlink{https://www.kaggle.com/c/avazu-ctr-prediction/data}{https://www.kaggle.com/c/avazu-ctr-prediction/data}} and KDD Cup 2012\footnote{\hyperlink{https://www.kaggle.com/c/kddcup2012-track2/data}{https://www.kaggle.com/c/kddcup2012-track2/data}}. All three datasets are pre-processed in the same fashion as AutoCTR~\cite{song2020towards}.

\subsection{Search Configuration}
We first demonstrate the detailed configuration of \textit{NASRec-Full} search space as follows:\
\begin{itemize}[noitemsep,leftmargin=*]
    \item \textbf{Connection Search Components.} We utilize $N=7$ blocks in our NASRec search space. This allows a fair comparison with recent NAS methods~\cite{song2020towards}. All choice blocks can arbitrarily connect to previous choice blocks or raw features.
    \item \textbf{Operator Search Components.} In each choice block, our search space contains 6 distinct building operators, including 4 dense building operators: FC, Gating, Sum, Dot-Product and 2 distinct sparse building operators: EFC and Attention.
    The dense-sparse merger option is fully explored. 
    \item \textbf{Dimension Search Components.} For each dense building operator, the dense output dimension can choose from \{16, 32, 64, 128, 256, 512, 768, 1024\}. For each sparse building operator, the sparse output dimension can be chosen from \{16, 32, 48, 64\}.
\end{itemize}
In \textit{NASRec-Small}, we employ the same settings except that we use only 2 dense building operators: FC, Dot-Product and 1 sparse building operator: EFC. 
Then, we illustrate some techniques on brewing the NASRec supernet, including the configuration of embedding, supernet warm-up, and supernet training settings.
\begin{itemize}[noitemsep,leftmargin=*]
    \item \textbf{Capped Embedding Table.} 
    We cap the maximum embedding table size to 0.5M during supernet training for search efficiency.
    During the final evaluation, we maintain the full embedding table to retrieve the best performance, i.e.,  a total of 540M parameters in DLRM~\cite{naumov2019deep} on Criteo to ensure a fair comparison.
    
    \item \textbf{Supernet Warm-up.}
    We observe that the supernet may collapse at initial training phases due to the varying sampled paths and uninitialized embedding layers.
    To mitigate the initial collapsing of supernet, we randomly sample the full supernet at the initial $1/5$ of the training steps, with a probability $p$ that linearly decays from 1 to 0. 
    This provides dimension warm-up, operator warm-up~\cite{bender2020can} and connection warm-up for the supernet with minimal impact on the quality of sampled paths.
    
    \item \textbf{Supernet Training Settings.} 
    We insert layer normalization~\cite{ba2016layer} into each building operator to stablize supernet training.
    Our choice of hyperparameters is robust over different NASRec search spaces and recommender system benchmarks. 
    We train the supernet for only 1 epoch with Adagrad optimizer, an initial learning rate of 0.12, a cosine learning rate schedule~\cite{loshchilov2016sgdr} on target recommender system benchmarks. 
\end{itemize}

Finally, we present the details of regularized evolution and model selection strategies over NASRec search spaces.
\begin{itemize}[noitemsep,leftmargin=*]
\item \textbf{Regularized Evolution.} Despite the large size of \textit{NASRec-Full} and \textit{NASRec-small}, we employ an efficient configuration of regularized evolution to seek the optimal subnets from supernet.
Specifically, we maintain a population of 128 architectures and run regularized evolution for 240 iterations. In each iteration, we first pick up the best architecture from 64 sampled architectures from the population as the parent architecture, and generate 8 child architectures to update the population.

\item \textbf{Model Selection.} We follow the evaluation protocols in AutoCTR~\cite{song2020towards} and split each target dataset into 3 sets: training (80\%), validation (10\%) and testing (10\%). During weight-sharing neural architecture search, we train the supernet on the training set and select the top-15 subnets on the validation set.
We train the top-15 models from scratch, and select the best subnet as the final architecture, namely, NASRecNet.
\end{itemize}

\subsection{Recommender System Benchmark Results}
We train NASRecNet from scratch on three classic recommender system benchmarks, and compare the performance of models that are crafted by NASRec on three general recommender system benchmarks. In Table~\ref{tab:ctr_results}, we report the evaluation results of our end-to-end NASRecNets and a random search baseline which randomly samples and trains models in our NASRec search space.

\noindent \textbf{State-of-the-art Performance.}
Even within an aggressively large \textit{NASRec-Full} search space, NASRecNets achieve record-breaking performance over hand-crafted CTR models~\cite{naumov2019deep,guo2017deepfm,lian2018xdeepfm} with minimal human priors as shown in Table \ref{tab:ctr_results}.
Compared with AutoInt~\cite{song2019autoint}, the hand-crafted model that fabricates feature interactions with delicate engineering efforts, NASRecNet achieves $\sim0.003$ Log Loss reduction on Criteo, $\sim0.007$ Log Loss reduction on Avazu, and $\sim0.003$ Log Loss reduction on KDD Cup 2012, with minimal human expertise and interventions.

Next, we compare NASRecNet to the more recent NAS-crafted models. 
Compared to AutoCTR~\cite{song2020towards}, NASRecNet achieves the state-of-the-art (SOTA) Log Loss and AUC on all three recommender system benchmarks.
With the same scale of search space as AutoCTR (i.e., \textit{NASRec-Small} search space), NASRecNet yields 0.001 Log Loss reduction on Criteo, 0.005 Log Loss reduction on Avazu, and 0.003 Log Loss reduction on KDD Cup 2012.
Compared to DNAS~\cite{krishna2021differentiable} and PROFIT~\cite{gao2021progressive} which only focuses on configuring part of the architectures, such as dense connectivity, NASRecNet achieves at least $\sim 0.002$ Log Loss reduction on Criteo, justifying the significance of full architecture search on recommender systems.

By extending NASRec to an extremely large \textit{NASRec-Full} search space, NASRecNet further improves its result on Avazu and outperforms PROFIT by $\sim 0.002$ AUC improvement with on-par Log Loss, justifying the design of \textit{NASRec-Full} with aggressively large cardinality and minimal human priors. On Criteo and KDD Cup 2012, NASRec maintains the edge in discovering state-of-the-art CTR models compared to existing NAS methods~\cite{song2020towards,gao2021progressive,krishna2021differentiable}.

\noindent \textbf{Efficient Search within a Versatile Search Space.}
Despite a larger NASRec search space that presents more challenges to fully explore, NASRec achieves at least 1.7$\times$ searching efficiency compared to state-of-the-art efficient NAS methods~\cite{song2020towards,gao2021progressive} with significant Log Loss improvement on all three benchmarks.
This is greatly attributed to the efficiency of Weight-Sharing NAS applied on heterogeneous operators and multi-modality data.

We observe that a compact \textit{NASRec-Small} search space produces strong random search baselines, while a larger \textit{NASRec-Full} search space has a weaker baseline.
This is because with limited search budget, it is more challenging to discover promising models within a large search space.
Yet, the scalable WS-NAS tackles the exploration of full \textit{NASRec-Full} search space thanks to the broad coverage of the supernet. With an effective Single-Operator Any-connection path sampling strategy, WS-NAS improves the quality of discovered models on Criteo, and discovers a better model on Avazu and KDD Cup 2012 compared to the NASRec-Small search space.

\subsection{Discussion}
In this section, we analyze the complexity of NASRecNet, and demonstrate the impact of our proposed techniques that mitigates ranking disorders and improve the quality of searched models.

\noindent \textbf{Model Complexity Analysis.}
We compare the model complexity of NASRecNets with SOTA hand-crafted and NAS models.
We collect all baselines from AutoCTR~\cite{song2020towards}, and compare performance versus the number of Floating-point Operations (FLOPs) in Table \ref{tab:model_complexity}. 

We profile all FLOPS of NASRecNets using FvCore~\cite{fvcore}.
Even without any FLOPs constraints, NASRecNets outperform existing arts in efficiency. Despite achieving lower Log Loss, NASRecNets achieve 8.5$\times$, 3.8$\times$, and 2.8$\times$ FLOPS reduction on Criteo, Avazu, and KDD Cup 2012 benchmarks.
One possible reason lies in the use of operator-balancing interaction modules: it projects the sparse inputs to a smaller dimension before carrying cross-term feature interaction. This leads to significantly lower computation costs, contributing compact yet high-performing recommender models.

\noindent \textbf{Effects of Path Sampling \& Fine-tuning.} We discussed the path sampling and fine-tuning techniques in Section \ref{sec:path_ft}, and demonstrate the empirical evaluation of these techniques on the quality of searched models in Table \ref{tab:effect_pt}. The results show that, (1) the importance of path sampling far outweigh the importance of fine-tuning in deciding the quality of searched models, and (2) a higher Kendall's $\tau$ that correctly ranks subnets in NASRec search space (i.e., Table \ref{tab:effect_pt}) indicates a consistent improvement on searched models.

\begin{table}[t]
    \caption{Model Complexity Analysis.}
    \vspace{-1.5em}
    \begin{center}
    \scalebox{0.75}{
    \begin{tabular}{|c|ccc|ccc|}
    \hline
         \multirow{2}{*}{\textbf{Method}} & \multicolumn{3}{|c|}{\textbf{Log Loss}} & \multicolumn{3}{|c|}{\textbf{FLOPS(M)}}  \\   
         & Criteo & Avazu & KDD & Criteo & Avazu & KDD \\
         \hline
         DLRM & 0.4436 & 0.3814 & 0.1523 & 26.92 & 18.29 & 25.84 \\
         DeepFM & 0.4432 & 0.3816 & 0.1529 & 22.74 & 22.50 & 21.66 \\
         AutoInt+ & 0.4427 & 0.3813 & 0.1523 & 18.33 & 17.49 & 14.88 \\
    \hline
    AutoCTR & 0.4413 & 0.3800 & 0.1520 & 12.31 & 7.12 & 3.02 \\
    NASRecNet @ \textit{NASRec-Small} & \textbf{0.4399} & 0.3747 & 0.1495 & 2.20 & 3.08 & 3.48 \\
    NASRecNet @ \textit{NASRec-Full} & 0.4408 & \textbf{0.3737} & \textbf{0.1491} & \textbf{1.45} & \textbf{1.87} & \textbf{1.09} \\
    \hline
    \end{tabular}        
    }
    \end{center}
    \vspace{-1.5em}
    \label{tab:model_complexity}
\end{table}

\begin{table}[t]
    \caption{Effects of different training techniques on NASRecNet, evaluated on Criteo.}
    \vspace{-1.5em}
    \begin{center}
    \scalebox{0.75}{

    \begin{tabular}{|c|c|c|}
    \hline
        \textbf{Method} & \textbf{Log Loss} & \textbf{FLOPS(M)}  \\
        \hline
        Baseline (Single-operator Any-connection + Fine-tuning) & \large 0.4408 & \large 1.45  \\
        Single-operator Single-connection + Fine-tuning & \large 0.4417 & \large 1.78 \\
        Any-operator Any-connection + Fine-tuning & \large 0.4413 & \large \large 2.04 \\
        Single-operator Any-connection, NO Fine-tuning &  \large 0.4410 & \large 3.62 \\
        \hline
    \end{tabular}
    }
    \end{center}
    \vspace{-2em}
    \label{tab:effect_pt}
\end{table}

\section{Conclusion}

In this paper, we propose NASRec, a new paradigm to fully enable \textbf{NAS} for \textbf{Rec}ommender systems via Weight Sharing Neural Architecture Search (WS-NAS) under data modality and architecture heterogeneity.
NASRec establishes a large supernet to represent the full architecture space, and incorporates versatile building operators and dense block connections to minimize human priors in automated architecture design for recommender systems.
NASRec identifies the scale and heterogeneity challenges of large-scale NASRec search space that compromises supernet and proposes a series of techniques to improve training efficiency 
and mitigate ranking disorder. 
Our crafted models, NASRecNet, achieve state-of-the-art performance on 3 popular recommender system benchmarks, demonstrate promising prospects on full architecture search space, and direct motivating research towards fully automated architecture fabrication with minimal human priors.

\noindent \textbf{Acknowledgement.}
Yiran Chen’s work is partially supported by the following grants:
NSF-2120333, NSF-2112562, NSF-1937435, NSF-2140247 and ARO W911NF-19-2-0107. Feng’s work is partially supported by the following grants: NSF CAREER-2048044 and
IIS-1838024. We also thank Maxim Naumov, Jeff Hwang and Colin
Taylor in Meta Platforms, Inc. for their kind help on this project.

\bibliographystyle{ACM-Reference-Format}
\bibliography{reference}

\newpage

\section{Supplementary Material}

In this section, we provide more details regarding NASRec, including: (1) the visualization and insight of searched architectures, (2) an evaluation of the best NASRecNet on Criteo Terabyte~\footnote{\hyperlink{https://ailab.criteo.com/download-criteo-1tb-click-logs-dataset/}{https://ailab.criteo.com/download-criteo-1tb-click-logs-dataset/}} to justify its performance on large-scale CTR prediction benchmarks, and (3) the details on subnet sampling and ranking. 

\subsection{Model Visualization}
We visualize the models searched within NASRec-Small/NASRec-Full search space on 3 different CTR benchmarks: Criteo, Avazu, and KDD.
Before presenting the searched architectures, we show the characteristics of each CTR benchmarks in Table \ref{tab:dataset_ch}.
\begin{table}[h]
\begin{center}
    \vspace{-1em}
    \caption{Statistics of different CTR benchmarks.}
    \vspace{-1em}
    \begin{tabular}{|c|c|c|c|}
        \hline
         textbf{Benchmark} & \# \textbf{Dense} & \# \textbf{Sparse} & \# \textbf{Samples (M)}  \\
         \hline
         Criteo & 13 & 26 & 45.84 \\
         Avazu & 0 & 23 & 40.42 \\
         KDD & 3 & 10 & 149.64  \\
         \hline
    \end{tabular}
    \label{tab:dataset_ch}    
\end{center}
\vspace{-1em}
\end{table}

Here, we observe that Criteo has the most number of dense (sparse) features, thus is the most complex and challenging benchmark. Avazu contains only dense features, thus requires less interactions between dense outputs in each choice block. KDD has the least number of features and the most data, making it a relatively easier benchmark to train and evaluate.

\begin{figure}[b]
    \vspace{-2em}
    \begin{center}
    \includegraphics[width=1.0\linewidth]{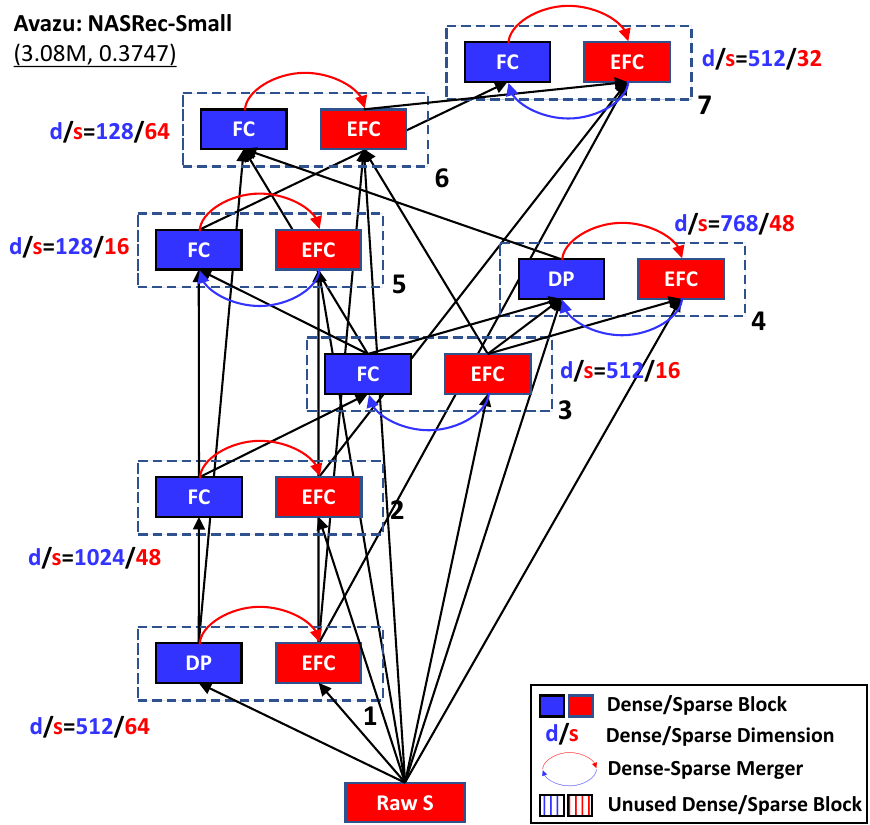}
    \caption{Best model discovered on Avazu @ NASRec-Small.}    
    \label{fig:avazu_best_nasrec_small}        
    \end{center}
\end{figure}

\begin{figure}[b]
\vspace{-2em}
\begin{center}
    \includegraphics[width=1.0\linewidth]{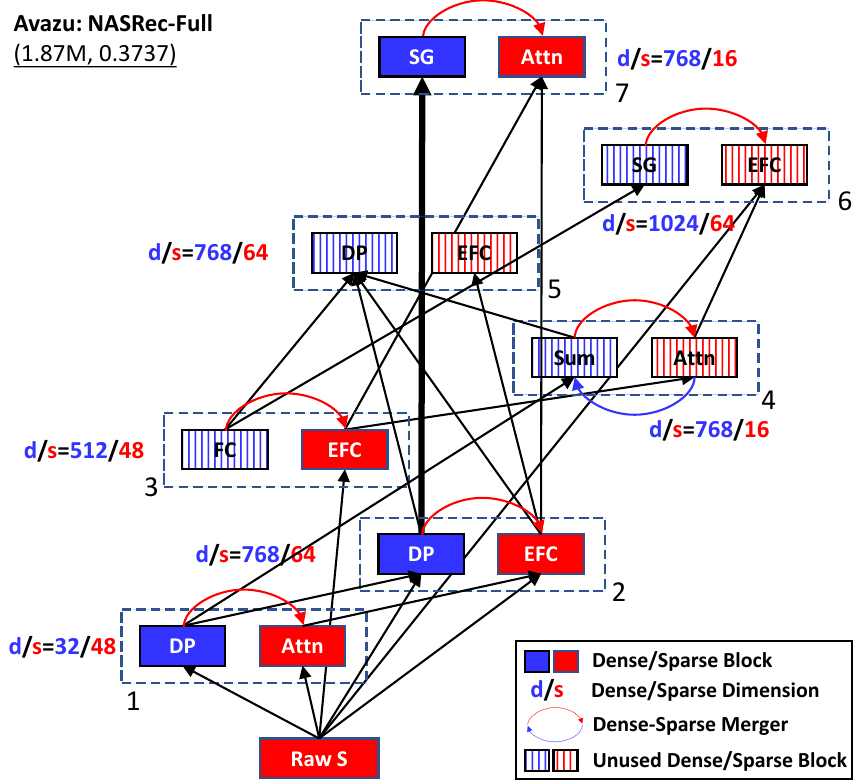}
    \caption{Best model discovered on Avazu @ NASRec-Full.}    
    \label{fig:avazu_best_nasrec_full}    
\end{center}
\end{figure}

\noindent \textbf{Avazu.} Figure \ref{fig:avazu_best_nasrec_small} and Figure \ref{fig:avazu_best_nasrec_full} depicts the detailed structures of best architecture within NASRec-Small/NASRec-Full search space. Here, a striped blue (red) block indicates an unused dense (sparse) block in the final architecture, and a bold connection indicates the same source input for a dense operator with two inputs (i.e., Sigmoid Gating and Sum).

As Avazu benchmark only contains sparse features, the interaction and extraction of dense representations are less important. For example, the best model within NASRec-Full search space only contains 1 operator (i.e., Sigmoid Gating) that solely processes dense representations, yet with more Dot-Product (DP) and Attention (Attn) blocks that interacts sparse representations.
Within NASRec-Small search space, dense representations are processed more frequently by FC layers after interacting with the sparse representations in the Dot-Product block. Yet, processing dense features require slightly more Fully-Connected blocks compared to the self-attention mechanism adopted in NASRec-Full search space.

\begin{figure}[b]  
    \vspace{-2em}
    \begin{center}
        \includegraphics[width=1.0\linewidth]{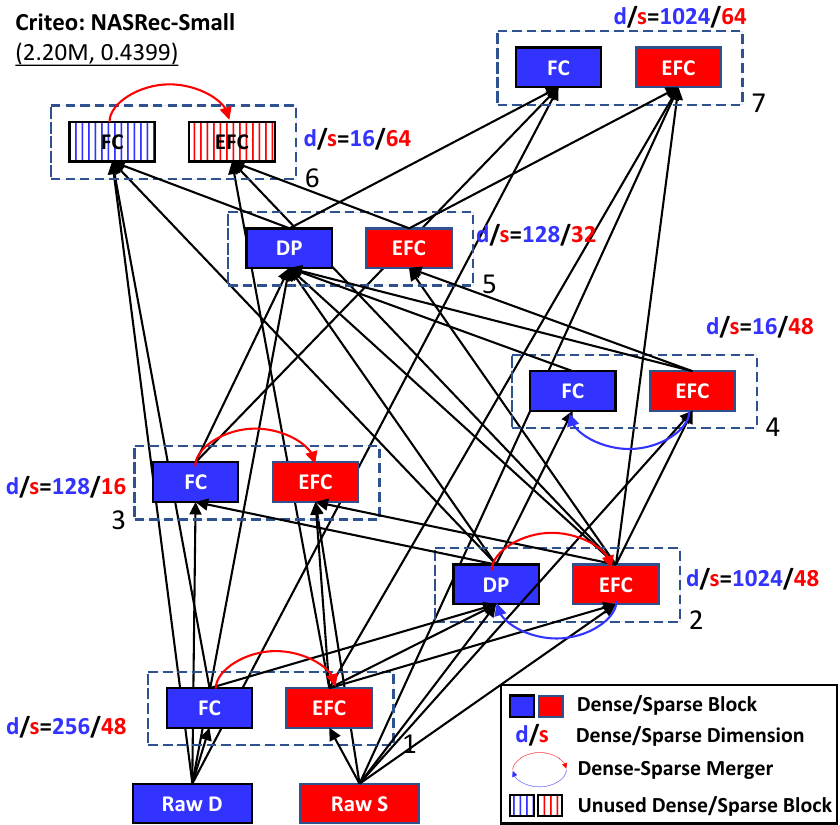}
        \caption{Best model discovered on Criteo @ NASRec-Small.}    
        \label{fig:criteo_best_nasrec_small}        
    \end{center}
\end{figure}

\begin{figure}[b]
    \vspace{-2em}
    \begin{center}
        \includegraphics[width=1.0\linewidth]{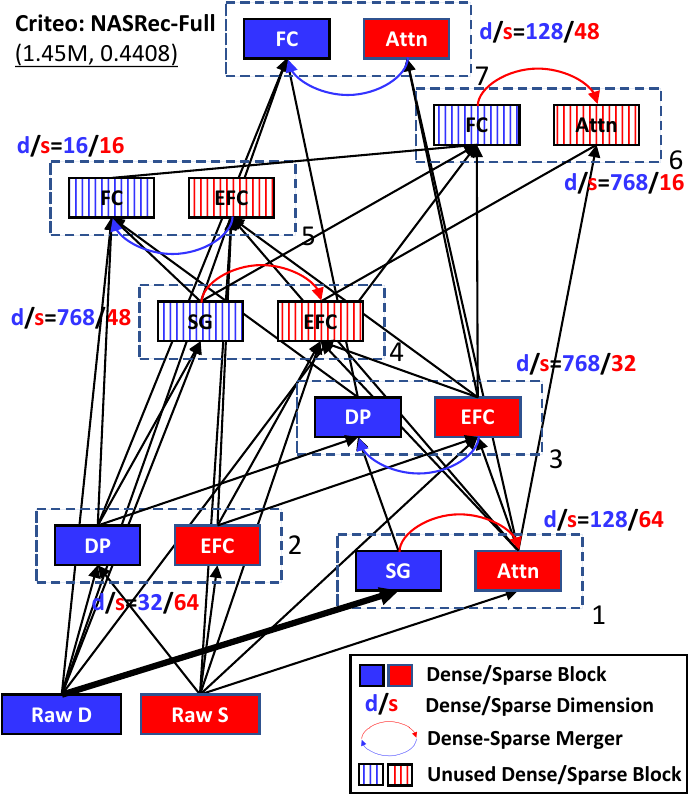}
        \caption{Best model discovered on Criteo @ NASRec-Full.}    
        \label{fig:criteo_best_nasrec_full}        
    \end{center}
\end{figure}

\noindent \textbf{Criteo.} Figure \ref{fig:criteo_best_nasrec_small} and Figure \ref{fig:criteo_best_nasrec_full} depicts the detailed structures of best architecture within NASRec-Small/NASRec-Full search space. Here, a striped blue (red) block indicates an unused dense (sparse) block in the final architecture, and a bold connection indicates the same source input for a dense operator with two inputs (i.e., Sigmoid Gating and Sum).

\begin{figure}[b]
    \vspace{-2em}
    \begin{center}
        \includegraphics[width=1.0\linewidth]{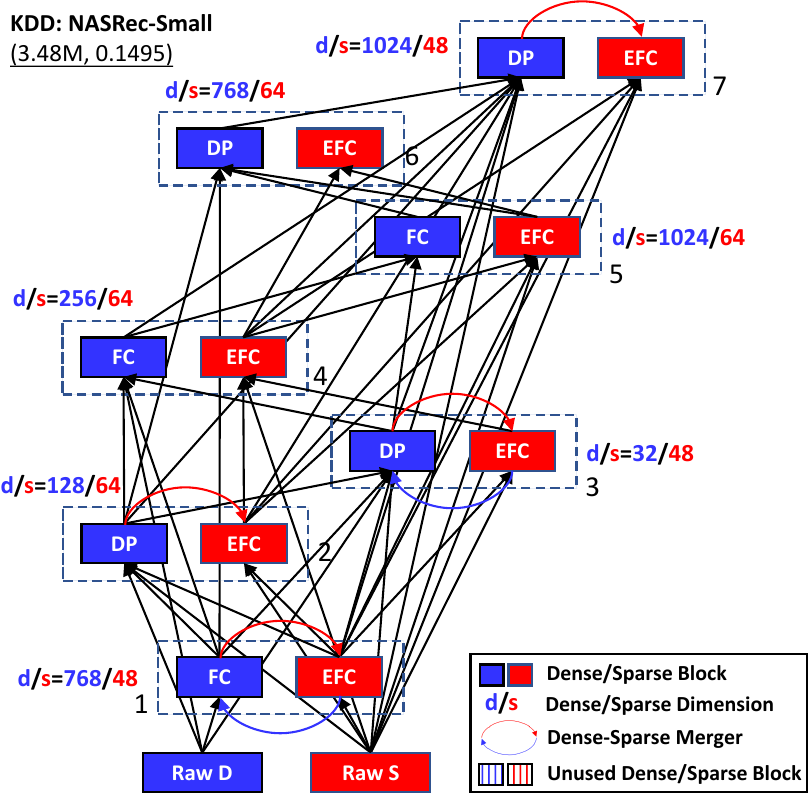}
        \caption{Best model discovered on KDD @ NASRec-Small.}
        \label{fig:kdd_best_nasrec_small}        
    \end{center}
\end{figure}

\begin{figure}[b]
    \vspace{-2em}
    \begin{center}
        \includegraphics[width=1.0\linewidth]{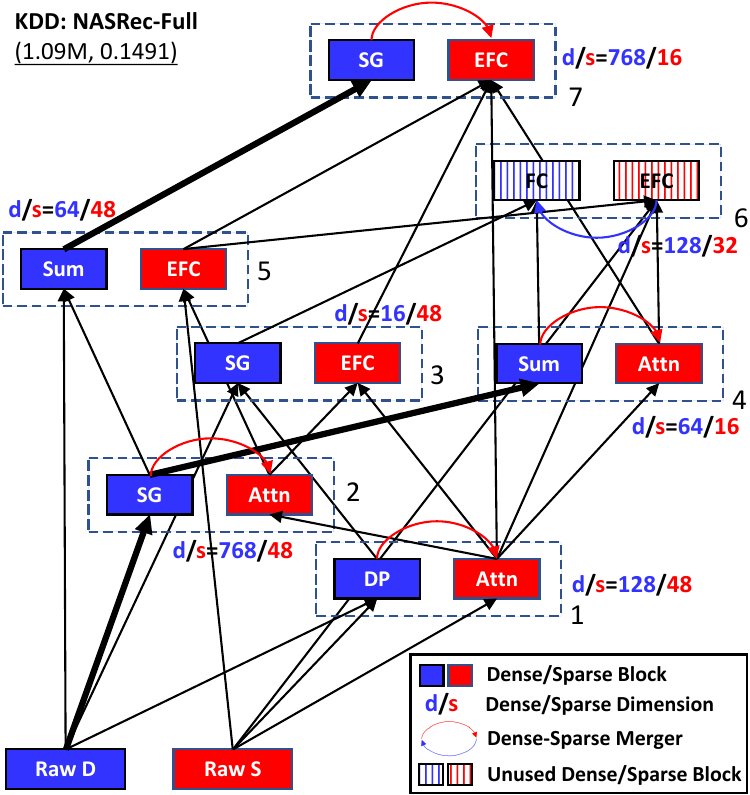}
        \caption{Best model discovered on KDD @ NASRec-Full.}    
        \label{fig:kdd_best_nasrec_full}        
    \end{center}
\end{figure}

Criteo contains the richest set of dense (sparse) features, thus is the most complex in architecture fabrication. we observe that dense connectivity is highly appreciated within both NASRec-Small and NASRec-Full search space, indicating that feature fusion is significantly impacting the log loss on a complex benchmarks. In addition, self-gating on raw dense features (i.e., block 1 @ NASRec-Full) is considered as an important motif in interacting features. Similar patterns can also be observed in the best architecture searched on KDD benchmarks.

Due to the complexity of Criteo and NASRec-Full search blocks, we notice that the best searched architecture does not use all of the 7 blocks in the search space. Some of the blocks are not utilized in the final architecture. For example, the best architecture searched within NASRec-Full contains only 4 valid blocks. We leave this as a future work to improve supernet training such that deeper architectures can be discovered in a more scalable fashion.

\noindent \textbf{KDD.} Figure \ref{fig:kdd_best_nasrec_small} and Figure \ref{fig:kdd_best_nasrec_full} depicts the detailed structures of best architecture within NASRec-Small/NASRec-Full search space. Here, a striped blue (red) block indicates an unused dense (sparse) block in the final architecture, and a bold connection indicates the same source input for a dense operator with two inputs (i.e., Sigmoid Gating and Sum). Similar to what we found on Criteo, the searched architecture within NASRec-Full has more building operators, yet less dense connectivity.

As KDD is a simpler benchmark with fewer dense (sparse) features, the searched architecture is simpler, especially within the NASRec search space. The similar self-gating on dense inputs still serve as an important motif in designing a better architecture.

In the end, we summarize our observations on three unique benchmarks as follows:
\begin{itemize}[noitemsep,leftmargin=*]
    \item \textbf{Benchmark Complexity Decides Architecture Complexity.} The choice of a benchmark decides the complexity of the final architecture. The more complex a benchmark is, the more complicated a searched model is in dense connectivity and operator heterogeneity.

    \item \textbf{Search Space Decides Connectivity.} On all three CTR benchmarks, the best architecture searched within NASRec-Full contains more operator heterogeneity and less dense connectivity. Yet, the reduced dense connectivity between different choice blocks help reduce FLOPs consumption of searched models, leading to less model complexity and better model efficiency. This also shows that the search for building operators may out-weigh the importance of the search for dense connectivity when crafting a efficient CTR model.

    \item \textbf{Attention Has a Huge Impact.} Attention blocks are rarely studied in existing literature on recommender systems. The architectures searched on NASRec-Full search space justifies the effectiveness of attention mechanism on aggregating dense (sparse) features. For example, the first block in the best searched architecture always adopt an attention layer to interact \textit{raw sparse inputs}. The stacking of attention blocks is also observed in searched architectures to demonstrate high-order interaction between dense (sparse) features.

    \item \textbf{Self-Gating Is a Useful Motif.} Self-gating indicates a pairwise gating operator with identical dense inputs. On both Criteo/KDD benchmark, self-gating is discovered to process \textit{raw dense inputs} and provide dense projections with a higher quality. On Avazu with no dense input features, self-gating is discovered to combine a higher-level dense representation for better prediction results.

\end{itemize}

\subsection{Evaluation on Criteo Terabyte}
Criteo Terabyte is a large-scale benchmark on CTR prediction, containing 1TB click logs within 24 days. Compared to the kaggle version of Criteo Kaggle \footnote{\hyperlink{https://www.kaggle.com/c/criteo-display-ad-challenge}{https://www.kaggle.com/c/criteo-display-ad-challenge}} which contains only 45.84M data, Criteo Terabyte contains $\sim$ 4B data in training and validation, thus has a significantly larger scale.

\begin{figure}[t]
    \includegraphics[width=1.0\linewidth]{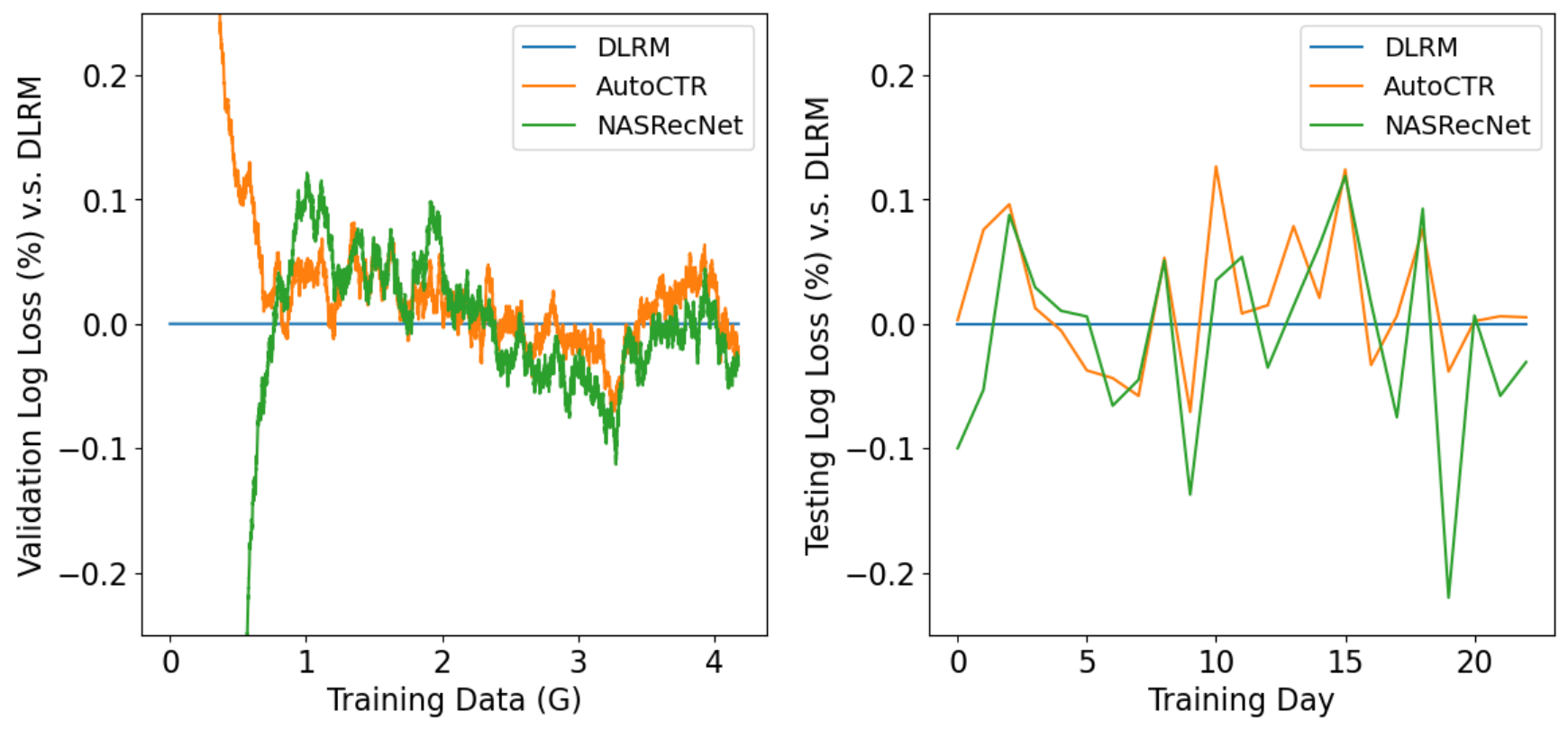}
    \vspace{-2em}
    \caption{Evaluation of best architecture on Criteo Terabyte.}    
    \vspace{-1em}
    \label{fig:criteo_terabyte_res}
\end{figure}

On Criteo Terabyte, we use the first 23 days of data as the training/validation data, and use the last day of data as testing data.
We evaluate DLRM, AutoCTR (i.e., the previous state-of-the-art) and NASRecNet models searched on the NASRec-Full search space. We plot the validation log loss on the \textbf{training} dataset and testing log loss on the testing dataset on Figure \ref{fig:criteo_terabyte_res}.
Compared to DLRM, AutoCTR shows on-par performance on testing dataset, yet NASRecNet achieves 0.03\% log loss reduction over DLRM baseline, showing better empirical results. However, as both AutoCTR and NASRecNet are crafted on the Criteo Kaggle dataset, they may not well suit the properties of a large-scale benchmark, such as data distribution shift. We leave the search and discovery for better architectures on large-scale benchmarks as future work.

\subsection{Subnet Sampling Details}
In Section 4, we sample 100 subnets within NASRec-Full search space on Criteo benchmark, with a more balanced and efficient setting on dimension search components: the dense output dimension can choose from \{32, 64, 128, 256, 512\}, and the sparse output dimension can choose from \{16, 32, 64\}.
All subnets are trained on the Criteo benchmark with a batch size of 1024 and a learning rate of 0.12.

\begin{figure}[h]
    \includegraphics[width=1.0\linewidth]{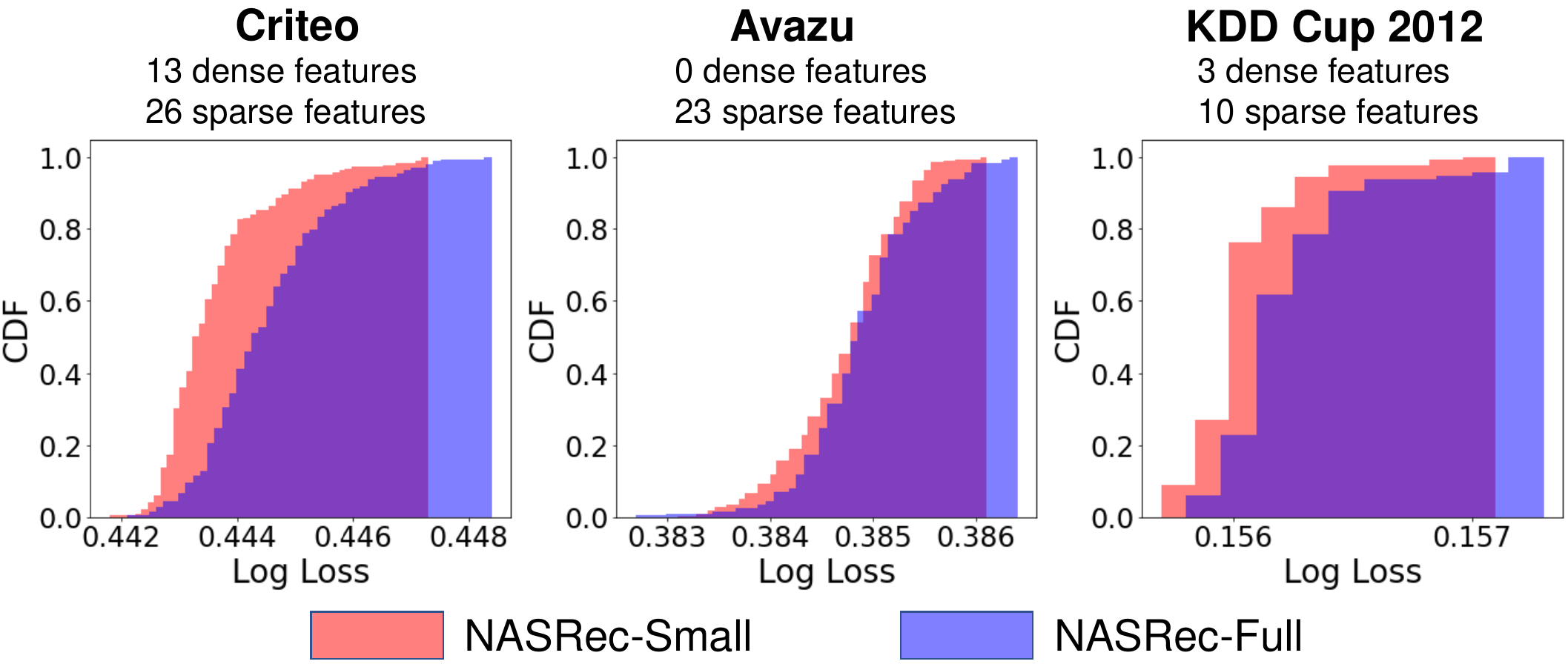}
    \vspace{-2em}
    \caption{CDF of log loss on CTR benchmarks.}    
    \vspace{-1em}
    \label{fig:subnet_cdf}
\end{figure}

We plot the CDF distribution of sampled subnets on all three benchmarks in Table \ref{fig:subnet_cdf}. For the top 50\% architectures evaluated on NASRec-Full supernet, we report a Kendall's $\tau$ of 0.24 for Criteo benchmark, showing a clear improvement on ranking top-performing architectures over the random search (0.0). In future work, we propose to establish a CTR benchmark for NAS to increase the statistical significance of evaluated ranking coefficients and better facilitate the research in accurately ranking different architectures. 

\end{document}